\documentclass[aip,jap,reprint]{revtex4-1}
\usepackage{amsfonts}
\usepackage{amsmath}
\usepackage{amssymb}
\usepackage{graphicx}
\usepackage{bm}

\begin{document}

\title{Electric field control of multiferroic domain wall motion}
\author{Hong-Bo Chen, Ye-Hua Liu, and You-Quan Li}
\affiliation{Zhejiang Institute of Modern Physics and Department of Physics, Zhejiang
University, Hangzhou 310027, P.R. China}

\begin{abstract}
The dynamics of a multiferroic domain wall in which an electric field can
couple to the magnetization via inhomogeneous magnetoelectric interaction is
investigated by the collective-coordinate framework. We show how the
electric field is capable of delaying the onset of the Walker breakdown of the domain wall motion,
leading to a significant enhancement of the maximum wall velocity. Moreover,
we show that in the stationary regime the chirality of the domain
wall can be efficiently reversed when the electric field is applied along
the direction of the magnetic field. These characteristics suggest that the
multiferroic domain wall may provide a new prospective means to design
faster and low-power-consumption domain wall devices.
\end{abstract}

\date{\today}
\maketitle

\section{Introduction}
Manipulation of magnetic properties by an external electric field has long
been a big challenge in the quest for novel spintronic devices. In
conventional ferroelectric or ferromagnetic materials, the controlled motion
of ferroic domain walls (DWs) is essential to achieve the desired
functionalities.\cite{Scott-RMP,Parkin-science-08} Usually, the motion of DW
in magnetic materials are driven by a magnetic field\cite%
{Schryer-JAP-1974,Wang1,Beach2005} or spin-polarized current.\cite%
{Berger-84,Tatara-prl-04} A key concept in the context of
DW motion is the so-called Walker breakdown,\cite{Schryer-JAP-1974} which
distinguishes the two regimes with high- and low-mobility and sets a limit
to the DW velocity. To achieve fast and efficient control of DW motion,
various attempts have been made to prevent this breakdown process, such as
applying a transverse field\cite{Allwood2008,Kunz2008,Glathe2008,Richter2010} or
considering a perpendicular magnetic anisotropy\cite{PMA1,PMA2,PMA3} and the
spin-orbit coupling effect.\cite{SOI-Tatara,SOI-DMI,SOI-SHE,SOI-exp1,SOI-exp2}%
 In this context, a way to manipulate the dynamics of the DWs by an
electric field, which is critical for developing low-power-consumption spintronic
devices, would be extremely appealing. Recently, it has been shown that
modulating the magnetic anisotropy by an applied electric field is possible,%
\cite{EM1,EM2} and thus will allow for the electric field control of DW
dynamics in ultrathin metallic ferromagnets.\cite%
{EDW1,EDW2,EDW3,EDW4,EDW5,EDW6,EDW7} Nevertheless, the search for
alternative schemes allowing fast and energy-efficient DW propagation is of
great relevance in advanced spintronics research.

Multiferroic materials,\cite{Cheong07-review} which exhibit simultaneously
ferroelectric and magnetic orders, may provide a promising arena to realize
electric control of magnetization and even for DW motion. Multiferroics
display a particularly rich variety of magnetoelectric (ME) cross-coupling
effects. An intriguing scenario of ME coupling is that spiral spin orders
can by themselves produce electric polarization, which is called the
spin-current mechanism,\cite{Nagaosa-05prl,Jia-prb07,Chen-JPC,Chen-APL} or equivalently the \textit{%
inhomogeneous magnetoelectric interaction}.\cite{Mostovoy-prl} Therefore, a
nonzero electric polarization can be induced not just in bulk but also
within local magnetic textures, like magnetic DWs and vortices. The
possibility of such a magnetoelectricity in ferromagnetic N\'{e}el walls has
been anticipated theoretically\cite{Mostovoy-prl,IME-1983} and recently
demonstrated experimentally.\cite{IME-exp1,IME-exp2}\ Especially, this ME\
coupling also enables the electric field couple to the magnetization with
its spatial gradients, which necessarily presents in metallic as well as
insulating ferromagnets.\cite{IME-SW1} The influence of this
coupling on the spin waves of the multiferroics has recently been explored in Ref.[%
\onlinecite{IME-SW2}]. However, the relevance of the electric field to the motion of
a multiferroic DW, even though it is crucial to future ME multiferroic
devices based on DW control, still remains unclear.

In this paper we identify the dynamical nature of a prototypical type of multiferroic DW, that is, a
magnetic DW simultaneously displaying an electric polarization. This is a
good basis for studying the dynamical properties of the multiferroic DWs, in
which the electric field can couple to the magnetization via the
inhomogeneous magnetoelectric interaction. We derive the equations of motion
for electric field controlled DW dynamics in such a multiferroic DW. We
report two main findings. The first one is that the magnetic DW velocity can be
considerably enhanced due to the delay of the occurrence of Walker breakdown
by an applied electric field. This electric-field-modulated higher DW speed
implies faster device operation, which is one of the main aim of the
conventional DW device applications. The second finding is that the electric
field can be used to control the switching of the DW chirality. This control
of the chirality could provide an additional degree of freedom, which can be
useful in future magnetoelectric logic devices.

The paper is organized as follows: In Sec. \ref{Sec:model} we present the
model for a multiferroic DW. We obtain the equations of motion for the
multiferroic DW dynamics using the collective coordinate description. In
Sec. \ref{Sec:discuss}, we discuss how the electric field influences the DW
velocity and the chirality switching. At the end, in Sec. \ref{Sec:summary}
we present a brief summary of the results obtained in this work.

\begin{figure}[tbp]
\centering\includegraphics[width=8cm]{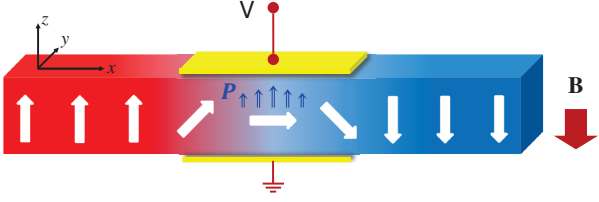}
\caption{(color
online) Schematic of the one-dimensional multiferroic DW structure consisting of a N\'{e}el-type magnetic DW which induces
an electric polarization (blue arrows) within the DW. A voltage is applied along the $z$ direction while the external magnetic field $\mathbf{B}$ is kept in $-\hat{z}$ direction. The white broad arrows denote the local spins in the wall. }%
\label{Fig:figure1}
\end{figure}

\section{Theoretical Model}

\label{Sec:model}

The system under consideration is schematically depicted in Fig. \ref%
{Fig:figure1}. We will focus on a case of one-dimensional insulating N\'{e}%
el-type DW when magnetic easy (hard) axis is taken to be along the $\hat{z}$
($\hat{y}$) direction with a voltage applied along the $z$ direction. The
theoretical model we employ consists of three distinct contributions, 
\begin{equation}
\mathcal{H}=\mathcal{H}_{\mathrm{S}}+\mathcal{H}_{\mathrm{Z}}+\mathcal{H}_{%
\mathrm{E}}.
\end{equation} The first contribution $\mathcal{H}_{\mathrm{S}}$ describes
the Hamiltonian of local spins in a magnetic DW, with an easy axis and a
hard axis, chosen as $\hat{z}$ and $\hat{y}$ directions, respectively. In
the continuum limit, this Hamiltonian takes the form,\cite{Tatara-review}
\begin{equation}
\mathcal{H}_{\mathrm{S}}=\int \frac{d^{3}x}{a^{3}}\left[ \frac{J}{2}\left(
\nabla \mathbf{S}\right) ^{2}-\frac{K}{2}(S_{z})^{2}+\frac{K_{\bot }}{2}%
(S_{y})^{2}\right] .  \label{Eq:Hs}
\end{equation}%
Here, $\mathbf{S}$ is the local spin vector, $a$ is the lattice constant, $J$
is the exchange coupling between local spins, while $K$ and $K_{\bot }$ are
anisotropy energies associated with the easy and hard axes of the spins,
respectively. Here we consider a homogeneous system, i.e., a system without
pinning potential. In terms of spherical coordinates ($\theta ,\phi $), the
energy functional in Eq. (\ref{Eq:Hs}) has a stationary DW as a classical
solution,
\begin{equation}
\theta (x)=2\arctan [e^{(x-X)/\lambda _{\mathrm{DW}}}],\ \ \ \phi (x)=\phi
_{0},  \label{DW-profile}
\end{equation}%
which gives the $x$-dependent spin configuration
\begin{equation}
\mathbf{S}=S(\sin \theta \cos \phi _{0},\sin \theta \sin \phi _{0},\cos
\theta ),
\end{equation}%
with $S$ is the magnitude of spin. In Eq. (\ref{DW-profile}), $X$ is the
position of the DW center, $\lambda _{\mathrm{DW}}=\sqrt{J/K}$ is the DW
width. $\phi _{0}$ is the tilting angle between spins at the DW center and
the easy plane, which is spatially homogeneous. In particular, if $\phi
_{0}=0$ or $\pi $, the domain wall considered above is called pure N\'{e}el
wall (as shown in Fig. \ref{Fig:figure1}) with opposite chirality: clockwise
(CW) or counterclockwise (CCW), respectively, while if $\phi _{0}=\pm \pi /2$ the wall
becomes a pure Bloch wall. Such a description of the DW assumes its
rigidity, that is, the DW can only move or rotate.

The second ingredient of our model is the Zeeman energy, which is given by%
\begin{equation}
\mathcal{H}_{\mathrm{Z}}=-\int d^{3}x\mathbf{M}\cdot \mathbf{B},
\label{Eq:Hz}
\end{equation}%
where $\mathbf{M}=-g\frac{\hbar }{a^{3}}\mathbf{S}$ is the magnetic moment
per unit volume, $g$(\textgreater$0$) is the gyromagnetic ratio and $\mathbf{%
B}$\ is a uniform external magnetic field pointing along the negative $z$
direction, i.e., $\mathbf{B}=-B\hat{z}$, $B>0$.

The third contribution $\mathcal{H}_{\mathrm{E}}$ models the coupling
between the electric field and the spin degrees of freedom of the multiferroic DW.
According to the inhomogeneous magnetoelectric interaction,\cite{Mostovoy-prl}%
 we take the following form%
\begin{equation}
\mathcal{H}_{\mathrm{E}}=-\mathbf{E}\cdot \mathbf{P},  \label{Eq:Hme}
\end{equation}%
with the electric polarization $\mathbf{P}$ induced within the DW given as%
\cite{Mostovoy-prl}%
\begin{equation}
\mathbf{P}=\gamma _{0}\int \frac{d^{3}x}{a^{3}}[\mathbf{S}(\nabla \cdot
\mathbf{S})-(\mathbf{S}\cdot \nabla )\mathbf{S}],  \label{Eq:Polar}
\end{equation}%
where $\gamma _{0}$ is the magnetoelectric coupling coefficient. One can
learn immediately from Eq. (\ref{Eq:Polar}) that inhomogeneous
magnetoelectric interaction induces electric polarization $\mathbf{P}$
within the N\'{e}el wall so that it is actually multiferroic.

To study the dynamics of a rigid planar\ DW, we employ a well-known
collective coordinate description.\cite{Tatara-review} In this approach, the
position $X$ and angle $\phi _{0}$ in Eq. (\ref{DW-profile}) of the wall is
regarded as time-dependent collective coordinates $\{X(t),\phi _{0}(t)\}$.
The chirality of the DW is determined by the tilt angle $\phi _{0}(t)$. The
domain wall is described by a Lagrangian of local spins given by%
\begin{equation}
\mathcal{L}=\int \frac{d^{3}x}{a^{3}}\hbar S\dot{\phi}_{0}(\cos \theta -1)-%
\mathcal{H}.  \label{Eq:Lagran1}
\end{equation}%
The first term represents the spin Berry phase. Inserting the DW ansatz (Eq.(%
\ref{DW-profile})) into Eq. (\ref{Eq:Lagran1}) and integrating the space
coordinate, the DW Lagrangian in terms of $X(t)$ and $\phi _{0}(t)$ can be
written as%
\begin{equation}
\mathcal{L}=-\frac{\hbar NS}{\lambda _{\mathrm{DW}}}\left( X\dot{\phi}%
+v_{\bot }\sin ^{2}\phi _{0}-gBX-\gamma E\cos \phi _{0}\right) .
\label{Eq:Lagran2}
\end{equation}%
Here, $N=2A\lambda _{\mathrm{DW}}/a^{3}$ is the number of spins in the wall
region with $A$ being the cross-sectional area of the system, $v_{\bot
}=\lambda _{\mathrm{DW}}K_{\bot }S/2\hbar $ and $\gamma =\pi S\gamma
_{0}/2\hbar $. To derive the equations of motion of a DW, one further needs
to introduce the dissipation function $\mathcal{W}$ to incorporate the
Gilbert damping, which is written as\cite{Tatara-review}%
\begin{equation}
\mathcal{W}=\int \frac{d^{3}x}{a^{3}}\frac{\hbar \alpha }{2S}\mathbf{\dot{S}}%
^{2}=\frac{\alpha \hbar NS}{2}\left[ \left( \dot{X}/\lambda _{\mathrm{DW}%
}\right) ^{2}+\dot{\phi}_{0}^{2}\right] ,  \label{Eq:dissp}
\end{equation}%
where $\alpha $ is the Gilbert damping parameter. We then utilize the
generalized Euler-Lagrangian equation\cite{Tatara-review}%
\begin{equation}
\frac{d}{dt}\frac{\delta \mathcal{L}}{\delta \dot{q}}-\frac{\delta \mathcal{L%
}}{\delta q}=-\frac{\delta \mathcal{W}}{\delta \dot{q}},  \label{Eq:Edissp}
\end{equation}%
where $q$ represents $X(t)$ and $\phi _{0}(t)$ and the last term describes
the energy dissipated. The equations of motion for the collective
coordinates, derived from the Eqs.(\ref{Eq:Lagran2}), (\ref{Eq:dissp}) and (%
\ref{Eq:Edissp}), are given as follows:
\begin{subequations}
\label{Eq:EM1}
\begin{eqnarray}
\frac{\dot{X}}{\lambda _{\mathrm{DW}}}-\alpha \dot{\phi}_{0} &=&\frac{%
v_{\bot }}{\lambda _{\mathrm{DW}}}\sin 2\phi _{0}+\frac{\gamma E}{\lambda _{%
\mathrm{DW}}}\sin \phi _{0},  \label{Eq:EM-1} \\
\dot{\phi}_{0}+\alpha \frac{\dot{X}}{\lambda _{\mathrm{DW}}} &=&gB.
\label{Eq:EM-2}
\end{eqnarray}%
\end{subequations}
These equations provide a basic description of the multiferroic DW dynamics
under the magnetic and electric fields. Thus the application of the electric
field on the DW introduces an additional spin torque proportional to $\sin
\phi _{0}$ into the equations of motion. This new term will act as a
chirality stabilizer, influencing significantly the DW dynamics. In what follows, we solve Eqs. (\ref{Eq:EM1}) numerically, and calculate the
values for $X$ and $\phi _{0}$ after a sufficiently long time. The average
terminal velocity of the DW is defined as $v_{\mathrm{DW}}=\langle \dot{X}%
\rangle $, where the angular brackets refer to a long-time average. To do
the numerical simulation, we take a fixed value for the Gilbert damping
parameter $\alpha =0.02$. The initial DW tilt angle is set to $\phi
_{0i}=\phi _{0}(t=0)=0$ throughout, so the initial chirality of the DW is
clockwise.

\begin{figure}
\centering\includegraphics[width=8.5cm]{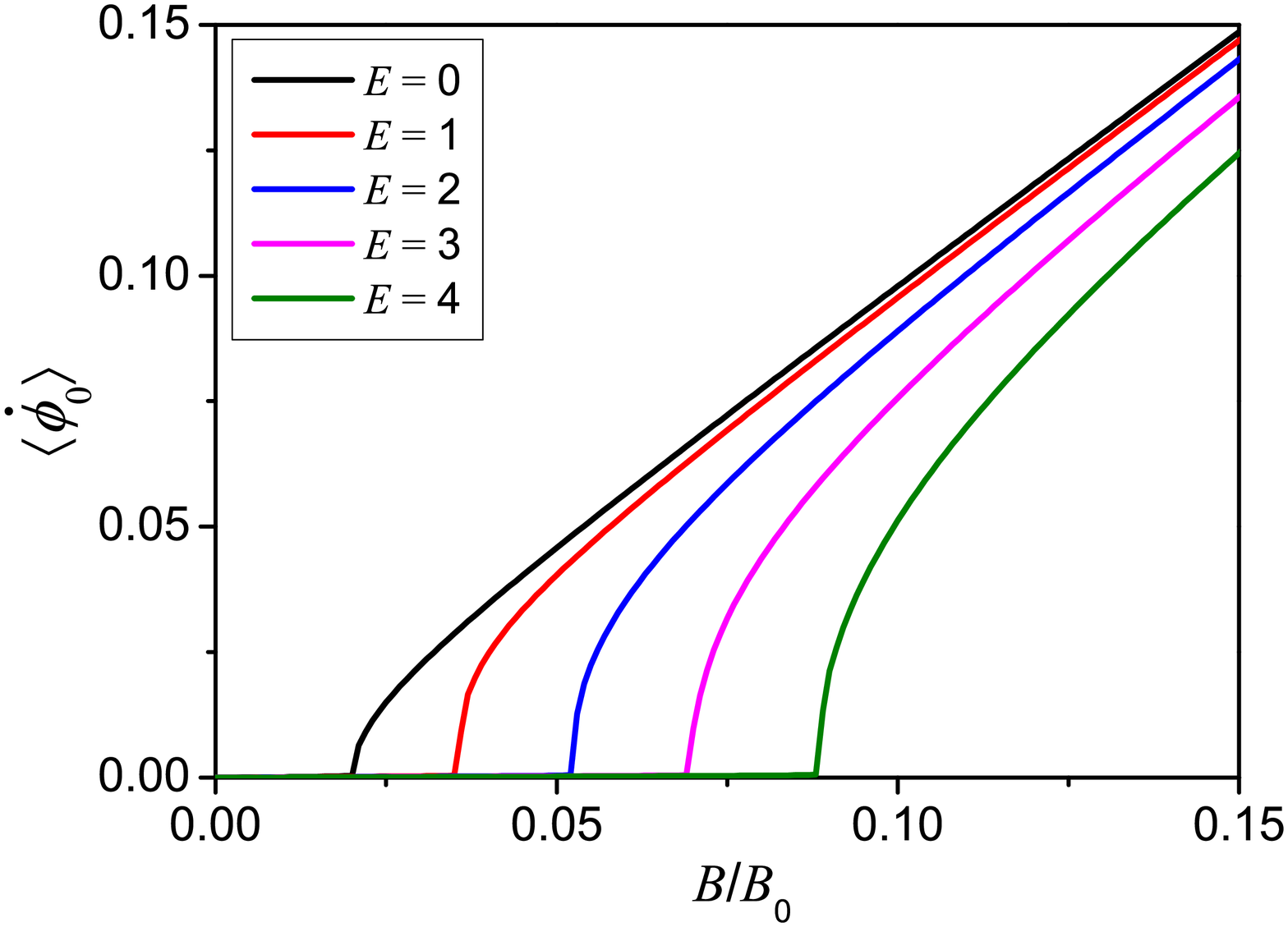}
\caption{(color
online) The behavior of the time-averaged precession velocity $\langle \dot{\phi}_{0}\rangle$ as a function of magnetic field $B$ at several applied electric field $E$.
$\langle \dot{\phi}_{0}\rangle$ is given in units of $K_{\perp}S/2\hbar$, the electric field $E$ is given in units of $E_{0}=v_{\perp}/\gamma$, and units $B_{0}=K_{\perp}S/2g\hbar$. The regime with $\langle \dot{\phi}_{0}\rangle=0$ represents the stationary regime of the DW motion.}
\label{Fig:figure2}
\end{figure}

\begin{figure}
\centering\includegraphics[width=8cm]{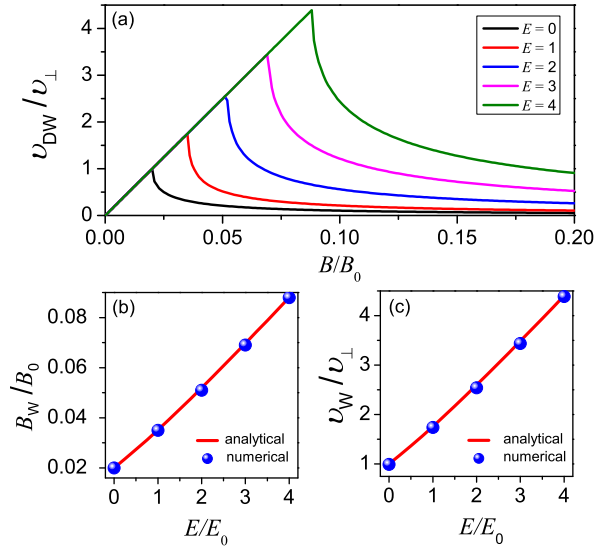}
\caption{(color
online) (a) The velocity of the DW $v_{\rm{DW}}$ as a function of
$B$ under the application of various $E$ applied along the $+\hat{z}$ direction.
$E$ is given in units of $E_{0}=v_{\perp}/\gamma$.
(b) and (c) are the dependence of the Walker
field $B_{\rm{W}}$ and the Walker velocity $v_{\rm{W}}$ with the electric field, respectively. Both show a quasi-linear increase with $E$.}
\label{Fig:figure3}
\end{figure}

\section{Results and Discussion}
\label{Sec:discuss}

\subsection{Electric field mediated DW velocity}

\begin{figure}[bp]
\centering\includegraphics[width=8.8cm]{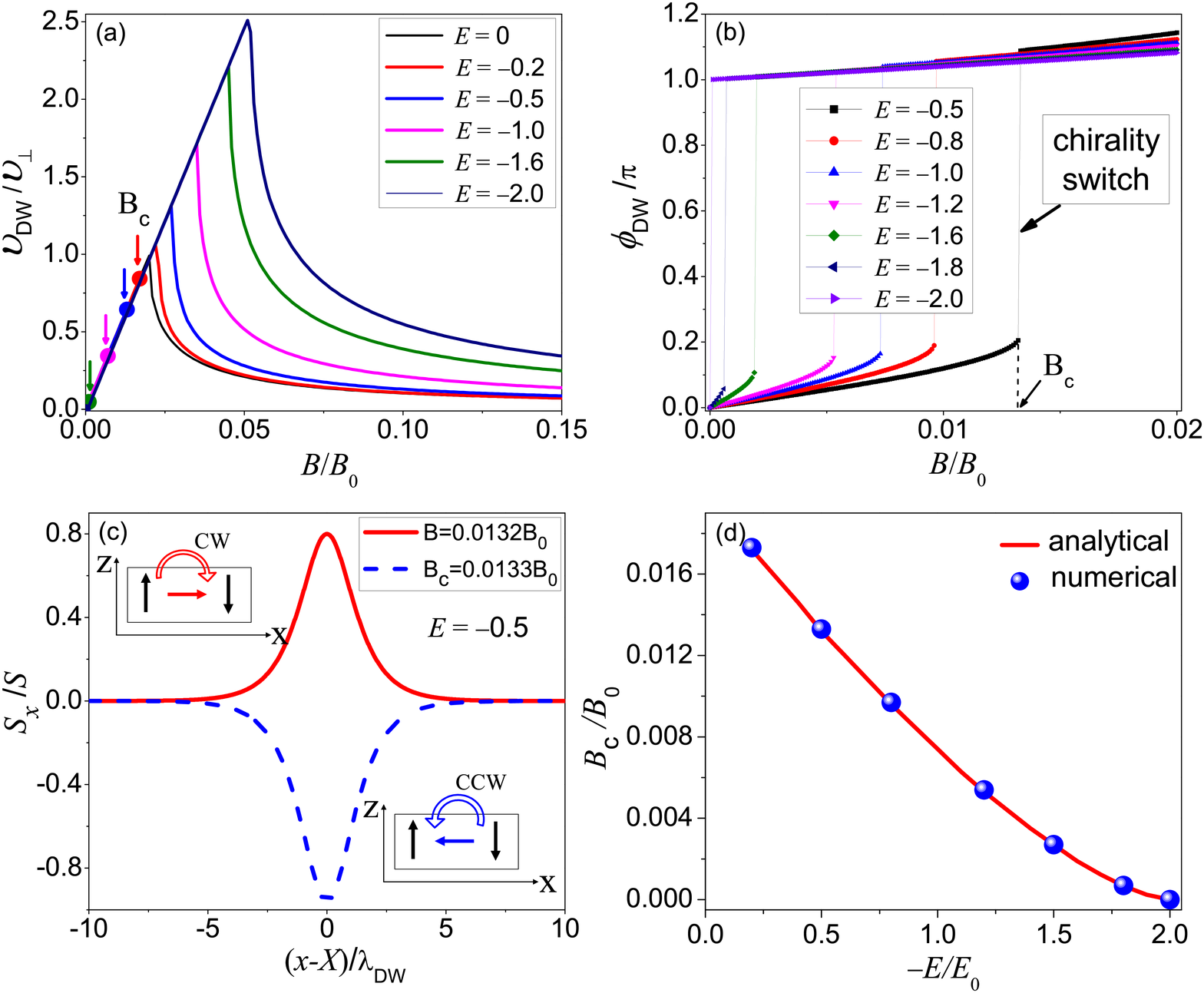}
\caption{(color
online) (a)$B$ dependence of the DW velocity $v_{\rm{DW}}$ for several choices of $E$ applied along the $-\hat{z}$ axis.
 The color dots mark the critical fields $B_{\rm{c}}$ indicating the chirality switching for each case.
(b) The terminal DW tilt angle $\phi_{\rm{DW}}$ as a function of $B$. The sudden jumps of $\phi_{\rm{DW}}$ denote the chirality switching occurring, from CW to CCW.
(c) Component of the spin along the $x$-axis (spin modulation direction) at the DW center. Below $B_{c}$, $S_{x}>0$ (red solid line) means CW  chirality, and above $B_{c}$, $S_{x}<0$ (blue dash line) means CCW  chirality. The applied $E=-0.5E_{0}$. (d) The dependence of $B_{\rm{c}}$ with the field strength $|E|$.}%
\label{Fig:figure4}
\end{figure}

Let us discuss how the electric field affects the field-driven DW motion
based on the equations of motion we derived previously. The numerical
simulation results of Eqs. (\ref{Eq:EM1}) are presented in Figs. \ref%
{Fig:figure2}-\ref{Fig:figure4}.
It is important to note that when the electric field is switched off, the
equations of motion in Eqs. (\ref{Eq:EM1}) are reduced to those of a DW
purely driven by a magnetic field, whose behaviours are well known.\cite%
{Tatara-review,Schryer-JAP-1974,Wang1} In that case, the DW motion is
characterized by the existence of two dynamic regimes, separated by a
threshold field called Walker field.\cite{Schryer-JAP-1974} That is, for an
external field smaller than the Walker field $B_{\mathrm{W}}=\alpha K_{\bot
}S/2g\hbar $, the DW moves with a constant velocity which increases linearly
with the external magnetic field up to $B_{\mathrm{W}}$. In this regime, the
DW chirality which describes the sense of rotation of the spins in the wall
is preserved during the motion. For fields $B>B_{\mathrm{W}}$, the Walker
breakdown occurs and the DW undergoes oscillatory motion, which makes the DW
velocity decrease rapidly. Such a behaviour was originally predicted by
Schryer and Walker\cite{Schryer-JAP-1974} and was observed experimentally
for example by Beach \textit{et al}.\cite{Beach2005}.

We first show in Fig. \ref{Fig:figure2} the time-averaged precession
velocity $\langle \dot{\phi}_{0}\rangle $, as a function of magnetic field $%
B $ for various values of electric fields applied along $+\hat{z}$ direction.
We find that $\langle \dot{\phi}_{0}\rangle =0$ up to a threshold applied
field, even when the external electric field is switched on. This zero
precession velocity means that the wall angle $\phi _{0}$ tilts out of the
plane until it reaches a certain angle. From then on, it no longer changes.
In the regime where $\langle \dot{\phi}_{0}\rangle =0$ the wall moves at a
constant velocity. As $\langle \dot{\phi}_{0}\rangle $ becomes finite, the
wall tilt angle $\phi _{0}$ starts precessing, causing an oscillatory motion
that slows down the domain wall. In Fig. \ref{Fig:figure2}, we clearly see
that the zero $\langle \dot{\phi}_{0}\rangle $ regime (stationary regime) is
significantly extended by the application of an electric field.

Figure \ref{Fig:figure3}(a) shows the time-averaged DW velocity $v_{\mathrm{%
DW}}$ as a function of $B$ for several applied electric
fields. When the electric field is switched on, the $v_{\mathrm{DW}}(B)$
curves show similar behavior to that of the conventional magnetic field
driven model. For each applied $E$, $v_{\mathrm{DW}}$ reaches
a maximum velocity, namely Walker velocity ($v_{\mathrm{W}}$). The corresponding threshold magnetic field is Walker field ($B_{\mathrm{W}}$),
and above $B_{\mathrm{W}}$ the $v_{\mathrm{DW}}$ drops abruptly. More
specifically, the Walker field $B_{\mathrm{W}}$ increases with $E$ and there
is no change of DW mobility. It seems that the presence of an electric field surely acts as
a chirality stabilizer and plays a pivotal role to delay the onset of the
Walker breakdown and allows for
higher attainable DW velocities. Figures \ref{Fig:figure3}%
(b) and (c) summarize the increase of both the Walker field $B_{\mathrm{W}}$
and the Walker velocity $v_{\mathrm{W}}$ with the magnitude of $E$. It is clearly shown that both $B_{\mathrm{W}}$ and $v_{\mathrm{W}}$
exhibit a nearly linear behavior, so we have an scenario where the maximum velocity of the wall is substantially enhanced by the application of an electric field. 

To elucidate the role of the electric field in the suppression of Walker
breakdown, we further examine analytically the DW dynamics for $B$ smaller
than $B_{\mathrm{W}}$, since in such a stationary regime $\partial \phi
_{0}/\partial t=0$ as $t\rightarrow \infty $. From Eqs. (\ref{Eq:EM1}), we
obtain
\begin{equation}
\frac{B}{\alpha B_{0}}=\sin 2\phi _{0}+\Delta \sin \phi _{0},  \label{Eq:Bw}
\end{equation}%
where $B_{0}=K_{\bot }S/2g\hbar $ and $\Delta =E/E_{0}$ with $E_{0}=v_{\bot
}/\gamma $. As long as this equation is satisfied, the DW will
propagate without oscillatory motion. The Walker field $B_{\mathrm{W}}$ is determined from the maximum
of the r.h.s. of Eq. (\ref{Eq:Bw}). Equation (\ref{Eq:Bw}) shows that $B_{%
\mathrm{W}}$ depends not only on the sign of $\Delta $ (or $E$) but also on
the initial tilt angle $\phi _{0i}(=\phi _{0}(t=0))$ (either $0$ or $\pi $).
The Walker field $B_{\mathrm{W}}$ is found to be associated with the tilt
angle $\phi _{0\mathrm{W}}=\arccos [(-\Delta +\sqrt{\Delta ^{2}+32})/8]$.
Then we can obtain the Walker field as $B_{\mathrm{W}}(\phi _{0i}=0,E>0)=B_{%
\mathrm{W}1}$ with
\begin{equation}
B_{\mathrm{W}1}=\frac{( 16-\Delta ^{2}+\left\vert \Delta \right\vert \tilde{%
\Delta}) ^{1/2}(3\left\vert \Delta \right\vert +\tilde{\Delta})\alpha B_{0}}{%
16\sqrt{2}},
\end{equation}%
where $\tilde{\Delta}=(\Delta ^{2}+32)^{1/2}$. For a large $E$ one has
simply $B_{\mathrm{W}}\sim \alpha E$, and further, the Walker velocity $v_{%
\mathrm{W}}\sim E$. The analytical results of $B_{\mathrm{W}}$ and $v_{%
\mathrm{W}}$ versus $E$ are shown as red solid lines in Figs. \ref%
{Fig:figure3}(b) and (c), respectively, which are both consistent with the
numerical simulations.

\subsection{Electric field induced DW chirality switching}

\label{Sec:discuss-b}

In this section, we investigate the effect of the applied electric field on
the chirality of DW. We show that a reliable control of the chirality
switching of a moving DW will be achieved by the application of an electric
field along the direction of the magnetic field. Each DW has two possible
chiralities: clockwise (CW) and counterclockwise (CCW). This DW chirality can be used
as an information unit.\cite{Parkin-science-08} Therefore, a controllable switching of the DW
chirality is desirable.

We now change the direction of the applied electric field to align the $-%
\hat{z}$ direction, i.e., parallel to the magnetic field. The numerical
results are shown in Fig. \ref{Fig:figure4}. Figure \ref{Fig:figure4}(a)
illustrates the $v_{\mathrm{DW}}(B)$ curves for various $E$. We can see that
a negative $E$ also results in a suppression of the Walker breakdown and an increase of the DW velocity, similar to the action of a positive $E$
discussed in Sec. \ref{Sec:discuss}. On the other hand, the chirality of
the DW is determined by the tilt angle $\phi _{0}(t)$. It should be noted
that $\phi _{0}(t)$ increases from the initial tilt angle $%
\phi _{0i}$ but eventually becomes saturated to a constant value in the
limit $t\rightarrow \infty $ below the Walker breakdown. Hence, controlling
the terminal tilt angle $\phi_{0} $ can be used to switch the chirality of the
DW. For a moving DW driven purely by a magnetic field, its initial chirality
is preserved below the Walker field,\cite{Tatara-review} and the chirality switching is
difficult to achieve in a controllable way. However, it can be shown that
this picture will not hold for multiferroic DW when $E$ is included. Figure %
\ref{Fig:figure4}(b) shows the terminal DW tilt angle $\phi _{\mathrm{DW}}$ (%
$=\phi _{0}(t\rightarrow \infty )$) as a function of the external magnetic
field $B$ in the presence of electric field $E$. We can see that below the
Walker field, the DW tilt angle $\phi _{\mathrm{DW}}$ initially increases
with $B$ from zero, and then suddenly jumps to a value larger than $\pi $ at a
critical field $B_{\mathrm{c}}$ for each $E$. Therefore, the switching of
the DW chirality from the initially clockwise to terminally counterclockwise
has clearly occurred at these critical magnetic fields. Fig. \ref{Fig:figure4}(c) further shows the $x$-component of the spin at the DW center for the applied fields $B$ near the critical field $B_{c}$. The applied electric field is chosen as $E=-0.5E_{0}$. The signs of the spin component along the $x$-axis indicate that below $B_{c}$ (i.e., $B=0.0132B_{0}$), the DW chirality is CW, when $B$ reaches up to $B_{c}$, the DW chirality switches to CCW. The positions of $B_{%
\mathrm{c}}$ extracted from the $\phi _{\mathrm{DW}}(B)$ curves are shown as
blue spheres in Fig. \ref{Fig:figure4}(d). It is shown that $B_{%
\mathrm{c}}$ decreases with $\left\vert E\right\vert $ down to zero at $%
\left\vert E\right\vert =2E_{0}$.

In order to better understand this remarkable chirality switching process, we take
the following analytical analyses. For a negative $E$, we can obtain from
Eq. (\ref{Eq:Bw}) the Walker field $B_{\mathrm{W}}(\phi _{0i}=0,E<0)=B_{%
\mathrm{W2}}$, with
\begin{equation}
B_{\mathrm{W2}}=\frac{(16-\Delta ^{2}-\left\vert \Delta \right\vert \tilde{%
\Delta})^{1/2}(-3\left\vert \Delta \right\vert +\tilde{\Delta})\alpha B_{0}}{%
16\sqrt{2}}.
\end{equation}%
For smaller negative $E$ we have $B_{\mathrm{W}%
}(\phi _{0i}=0,E<0)\sim -\alpha E$. Interestingly, if setting the initial tilt angle $\phi _{0i}=\pi$ (i.e., the initial chirality is CCW), we can
also obtain the Walker field as $B_{\mathrm{W}}(\phi _{0i}=\pi ,E<0)=B_{%
\mathrm{W1}}$. In this case, the terminal tilt angle $\phi _{\mathrm{DW}}$ is larger than
$\pi $. Moreover, it is easy to see that for $E<0$, $B_{\mathrm{W}}(\phi _{0i}=0,E<0)$ is smaller than $%
B_{\mathrm{W}}(\phi _{0i}=\pi ,E<0)$. Such a difference between the two
Walker fields $B_{\mathrm{W}}(\phi _{0i}=0,E<0)$ and $B_{\mathrm{W}}(\phi
_{0i}=\pi ,E<0)$ enables the chirality of a moving DW to be switched. That
is, when $B$ increases beyond the first Walker field $B_{\mathrm{W}}(\phi
_{0i}=0,E<0)$, the Walker breakdown process does not occur until $B$ further
reaches the higher Walker field $B_{\mathrm{W}}(\phi _{0i}=\pi ,E<0)$. In
the meantime, the initial tilt angle $\phi _{0i}$ switches from $0$ to $\pi $
at the first threshold field $B_{\mathrm{W}}(\phi _{0i}=0,E<0)$ and thus the
DW chirality can be switched. The $E$ dependence of $B_{\mathrm{W}%
}(\phi _{0i}=0,E<0)$ is shown in Fig. \ref{Fig:figure4}(d) as
red solid line, which is in excellent agreement with the numerical results $%
B_{\mathrm{c}}$ extracted from the $\phi _{\mathrm{DW}}(B)$ curves. However,
for $E>0$, $B_{\mathrm{W}}(\phi _{0i}=0,E>0)>B_{\mathrm{W}}(\phi _{0i}=\pi
,E>0)$, we have only one threshold field $B_{\mathrm{W}}(\phi
_{0i}=0,E>0) $ and hence there is no DW chirality switching process. The switching of the DW chirality with
the application of electric field can be read by a magnetic field sensor
since the stray field near a DW depends on its chirality. It
should be noted that the controlled chirality switching of a moving DW can
also be achieved by applying an oblique magnetic field as proposed by Seo
\textit{et al}. \cite{HWLee-chirality}. Nevertheless, we here offer a more
efficient alternative to flip the chirality of the DW with the help of an
external electric field.

\section{Summary}

\label{Sec:summary}

We considered multiferroic DW systems that exhibit both a coexistence and a
coupling of electric polarization and a magnetic DW. The effects of electric
field on the DW dynamics via the inhomogeneous magnetoelectric interaction have
been investigated. We have revealed the dynamical nature of a multiferroic
DW and demonstrated the efficiency of an electric field control of magnetic
DW motion. In particular, we showed that the electric field can achieve not
only a nearly linear enhancement of the maximum wall velocity, but also a
controllable switching of DW chirality. This control of the motion of the multiferroic DWs via electric fields can be useful for designing low-power and high-speed DW-based magnetoelectric memory and logic devices.\\[8pt]

\section*{Acknowledgements}

This work was supported by NSF-China (Grant Nos. 11074216 and 11274272), and
the Fundamental Research Funds for the Central Universities in China.


\begin{thebibliography}{99}
\bibitem{Scott-RMP} G. Catalan, J. Seidel, R. Ramesh, and J. F. Scott, Rev. Mod.
Phys. \textbf{84}, 119 (2012).

\bibitem{Parkin-science-08} D. A. Allwood, G. Xiong, C. C. Faulkner, D.
Atkinson, D. Petit, and R. P. Cowburn, Science \textbf{309}, 1688 (2005); S.
S. P. Parkin, M. Hayashi, and L. Thomas, Science \textbf{320}, 190 (2008).

\bibitem{Schryer-JAP-1974} N. L. Schryer and L. R. Walker, J. Appl. Phys.
\textbf{45}, 5406 (1974).

\bibitem{Wang1} X. R. Wang, P. Yan, and J. Lu, Europhys. Lett. \textbf{86},
67001 (2009); B. Hu and X. R. Wang, Phys. Rev. Lett. \textbf{111}, 027205
(2013).


\bibitem{Beach2005} G. S. D. Beach, C. Nistor, C. Knutson, M. Tsoi, and J.
L. Erskine, Nature Mater. \textbf{4,} 741 (2005).

\bibitem{Berger-84} L. Berger, J. Appl. Phys. \textbf{55}, 1954 (1984).

\bibitem{Tatara-prl-04} G. Tatara and H. Kohno, Phys. Rev. Lett. \textbf{92}%
, 086601 (2004); S. Zhang and Z. Li, Phys. Rev. Lett. \textbf{93}, 127204
(2004); A. Thiaville, Y. Nakatani, J. Miltat, and Y.
Suzuki, Europhys. Lett. \textbf{69}, 990 (2005).

%

\bibitem{Allwood2008} M. T. Bryan, T. Schrefl, D. Atkinson, and D. A.
Allwood, J. Appl. Phys. \textbf{103}, 073906 (2008).

\bibitem{Glathe2008} S. Glathe, I. Berkov, T. Mikolajick, and R. Mattheis,
Appl. Phys. Lett. \textbf{93}, 162505 (2008).

\bibitem{Kunz2008} A. Kunz and S. C. Reiff, J. Appl. Phys. \textbf{103},
07D903 (2008).

\bibitem{Richter2010} K. Richter, R. Varga, G. A. Badini-Confalonieri, and
M. V\'{a}zquez, Appl. Phys. Lett. \textbf{96}, 182507 (2010).

\bibitem{PMA1} J.-Y. Lee, K.-S. Lee, and S.-K. Kim, Appl. Phys. Lett.
\textbf{91}, 122513 (2007).

\bibitem{PMA2} S.-W. Jung, W. Kim, T.-D. Lee, K.-J. Lee, and H.-W. Lee,
Appl. Phys. Lett. \textbf{92}, 202508 (2008).

\bibitem{PMA3} S. Emori and G. S. D. Beach, Appl. Phys. Lett. \textbf{98},
132508 (2011).

\bibitem{SOI-Tatara} K. Obata and G. Tatara, Phys. Rev. B \textbf{77},
214429 (2008).

\bibitem{SOI-DMI} O. A. Tretiakov and A. Abanov, Phys. Rev. Lett. \textbf{105%
}, 157201 (2010); A. Thiaville, S. Rohart, E. Jue, V. Cros, and A. Fert,
Europhys. Lett. \textbf{100}, 57002 (2012).

\bibitem{SOI-SHE} K.-W. Kim, S.-M. Seo, J. Ryu, K.-J. Lee, and
H.-W. Lee, Phys. Rev. B \textbf{85}, 180404(R) (2012); J. Ryu, K.-J. Lee, and H.-W. Lee, Appl. Phys. Lett.
\textbf{102}, 172404 (2013).

\bibitem{SOI-exp1} K. S. Ryu, L. Thomas, S. H. Yang, and S. S. P. Parkin,
Nature Nanotech. \textbf{8}, 527 (2013).

\bibitem{SOI-exp2} S. Emori, U. Bauer, S. M. Ahn, E. Martinez, and G. S. D.
Beach, Nature Mater. \textbf{12}, 611 (2013).

\bibitem{EM1} M. Weisheit, S. Fahler, A. Marty, Y. Souche, C. Poinsignon,
and D. Givord, Science \textbf{315}, 349 (2007).

\bibitem{EM2} M. Endo, S. Kanai, S. Ikeda, F. Matsukura, and H. Ohno, Appl.
Phys. Lett. \textbf{96}, 212503 (2010).

\bibitem{EDW1} U. Bauer, S. Emori, and G. S. D. Beach, Appl. Phys. Lett.
\textbf{100}, 192408 (2012).

\bibitem{EDW2} A. J. Schellekens, A. van den Brink, J. H. Franken, H. J. M.
Swagten, and B. Koopmans, Nature Commun. \textbf{3}, 847 (2012).

\bibitem{EDW3} D. Chiba, M. Kawaguchi, S. Fukami, N. Ishiwata, K. Shimamura,
K. Kobayashi, and T. Ono, Nature Commun. \textbf{3}, 888 (2012).

\bibitem{EDW4} U. Bauer, S. Emori, and G. S. D. Beach, Appl. Phys. Lett.
\textbf{101}, 172403 (2012).

\bibitem{EDW5} A. Bernand-Mantel, L. Herrera-Diez, L. Ranno, S. Pizzini, J.
Vogel, D. Givord, S. Auffret, O. Boulle, I. M. Miron, and G. Gaudin, Appl.
Phys. Lett. \textbf{102}, 122406 (2013).

\bibitem{EDW6} J. H. Franken, Y. Yin, A. J. Schellekens, A. van den Brink,
H. J. M. Swagten, and B. Koopmans, Appl. Phys. Lett. \textbf{103}, 102411
(2013).

\bibitem{EDW7} U. Bauer, S. Emori, and G. S. D. Beach, Nature Nanotech.
\textbf{8}, 411 (2013).

\bibitem{Cheong07-review} S.-W. Cheong and M. Mostovoy, Nature Mater.
\textbf{6}, 13 (2007).

\bibitem{Nagaosa-05prl} H. Katsura, N. Nagaosa, and A. V. Balatsky, Phys.
Rev. Lett. \textbf{95}, 057205 (2005).

\bibitem{Jia-prb07}C. Jia, S. Onoda, N. Nagaosa, and J.
H. Han, Phys. Rev. B \textbf{74}, 224444 (2006); C. Jia, S. Onoda, N. Nagaosa, and J. H. Han, Phys. Rev. B \textbf{76}, 144424 (2007).

\bibitem{Chen-JPC} H. B. Chen, Y. Zhou, and Y. Q. Li, J. Phys.: Condens.
Matter \textbf{25}, 286004 (2013).

\bibitem{Chen-APL} H. B. Chen and Y. Q. Li, Appl. Phys. Lett. \textbf{102},
252906 (2013).

\bibitem{Mostovoy-prl} M. Mostovoy, Phys. Rev. Lett. \textbf{96}, 067601
(2006).

\bibitem{IME-1983} V. G. Bar'yakhtar, V. A. Lvov, and D. A. Yablonskii, JETP
Lett. \textbf{37, }673 (1983).

\bibitem{IME-exp1} A. S. Logginov, G. A. Meshkov, A. V. Nikolaev, E. P.
Nikolaeva, A. P. Pyatakov, and A. K. Zvezdin, Appl. Phys. Lett. \textbf{93},
182510 (2008).

\bibitem{IME-exp2} A. P. Pyatakov, D. A. Sechin, A. S. Sergeev, A. V.
Nikolaev, E. P. Nikolaeva, A. S. Logginov, and A. K. Zvezdin, Europhys.
Lett. \textbf{93}, 17001 (2011).

\bibitem{IME-SW1} I. E. Dzyaloshinskii, Europhys. Lett. \textbf{83}, 67001
(2008).

\bibitem{IME-SW2} D. L. Mills and I. E. Dzyaloshinskii, Phys. Rev. B \textbf{%
78}, 184422 (2008).

\bibitem{Tatara-review} G. Tatara, H. Kohno and J. Shibata, Phys. Rep.
\textbf{468}, 213 (2008).

\bibitem{HWLee-chirality} S.-M. Seo, K.-J. Lee, S.-W. Jung, and H.-W. Lee,
Appl. Phys. Lett. \textbf{97}, 032507 (2010).
\end{thebibliography}
\end{document}